\documentclass[aps,pra,twocolumn,superscriptaddress]{revtex4-2}
\usepackage{graphicx}
\usepackage{url}
\usepackage{amsmath,amsthm,amssymb,physics,color,enumerate,tabularx,makecell,mathrsfs,float,bbm,stackrel,mathtools,multirow,bm}
\usepackage{natbib}
\usepackage[colorlinks=true,%
bookmarks=false,%
linkcolor=blue,%
urlcolor=blue,%
citecolor=blue,%
breaklinks]{hyperref}

\usepackage[dvipsnames]{xcolor}

\begin{document}

	\title{Charging a quantum battery in a non-Markovian environment: a collisional model approach}

	\author{Daniele Morrone}
\email{daniele.morrone@unimi.it}
\affiliation{Quantum Technology Lab, Dipartimento di Fisica {\it Aldo Pontremoli}, Universit\`{a} degli Studi di Milano, I-20133 Milano, Italy}
	\author{Matteo A. C. Rossi}%
\affiliation{InstituteQ - the Finnish Quantum Institute, Aalto University, Finland}
\affiliation{QTF Centre of Excellence, Department of Applied Physics, Aalto University, FI-00076 Aalto, Finland}
\affiliation{Algorithmiq Ltd., Kanavakatu 3C, FI-00160 Helsinki, Finland}
	
	\author{Andrea Smirne}
	\affiliation{Quantum Technology Lab, Dipartimento di Fisica {\it Aldo Pontremoli}, Universit\`{a} degli Studi di Milano, I-20133 Milano, Italy}
        \affiliation{Istituto Nazionale di Fisica Nucleare, Sezione di Milano, via Celoria 16, I-20133 Milano, Italy}
	\author{Marco G. Genoni}
	\email{marco.genoni@fisica.unimi.it}
\affiliation{Quantum Technology Lab, Dipartimento di Fisica {\it Aldo Pontremoli}, Universit\`{a} degli Studi di Milano, I-20133 Milano, Italy}

\date{\today}
	
\begin{abstract}
We study the effect of non-Markovianity in the charging process of an open-system quantum battery. We employ a collisional model framework, where the environment is described by a discrete set of ancillary systems and memory effects in the dynamics can be introduced by allowing these ancillas to interact. We study in detail the behaviour of the steady-state ergotropy and the
impact of the information backflow to the system on the different features characterizing the charging process. Remarkably, we find that there is a maximum value of the ergotropy achievable: this value can be obtained either in the presence of memoryless environment, but only in the large-loss limit, as derived in [D. Farina et al., Phys. Rev. B 99, 035421 (2019)], or in the presence of an environment with memory also beyond the large-loss limit. In general, we show that the presence of an environment with memory allows us to generate steady-state ergotropy near to its maximum value for a much larger region in the parameter space and thus potentially in a shorter time. Relying on the geometrical measure of non-Markovianity, we show that
in both the cases of an environment with and without memory the ergotropy maximum is obtained when the 
non-Markovianity of the dynamics of the battery is zero, possibly as the result of a non-trivial interplay between the memory effects induced by, respectively, the environment and the charger connected to the battery.
\end{abstract}

\maketitle
	
	
\section{\label{Introduction}Introduction}
A battery is a device meant to act as an energy (work) reservoir, where energy is injected during the charging process and later discharged into a consumption hub. The study of the performances of quantum batteries (QB), that is batteries whose energy charging and discharging processes are based on the laws of quantum mechanics, has both fundamental and technological motivations.

Since the seminal work of Alicki and Fannes \cite{alickiEntanglementBoostExtractable2013}, where the concept of quantum battery has first been introduced, various scholars explored this increasingly vast field, investigating the role of quantum resources in the charging of this kind of devices, their charging power bounds and the development of optimal charging protocols \cite{campaioliEnhancingChargingPower2017,binderQuantacellPowerfulCharging2015,ferraroHighPowerCollectiveCharging2018,campaioliQuantumBatteriesReview2018,gyhmQuantumChargingAdvantage2022}. Theoretical frameworks applied to the study of QBs are numerous, including collections of qubits, spin chains and harmonic oscillators \cite{campaioliQuantumBatteriesReview2018,Andolina_2018,rosaUltrastableChargingFastscrambling2020,rossiniQuantumAdvantageCharging2020}, covering for an equally large amount of experimental platforms available to implement QBs. As of recently, a Dicke-model QB has been realized through a cavity \cite{quachSuperabsorptionOrganicMicrocavity2022} and another has been implemented through superconducting qubits \cite{huOptimalChargingSuperconducting2021}.

To guarantee that a real-life implementation of QBs actually works despite the unavoidable interaction with the environment, it is important to include this interaction, and any dissipative effect it might entail, in the theoretical description. Moreover, the dissipation and decoherence brought by the interaction lead quantum batteries to a stationary state and, as a consequence, the stored energy to a stationary value, at variance with unitary charging protocols, where the energy has an oscillatory behaviour. For these reasons, much attention has been given to the study of QBs in an open-system setting, sometimes being referred to as open quantum batteries (OQBs).

Within this framework, there is interest not only in studying the effect of the environment on QBs \cite{farinaChargermediatedEnergyTransfer2019,zakavatiBoundsChargingPower2021,carregaDissipativeDynamicsOpen2020,kaminNonMarkovianEffectsCharging2020}, but also in developing open-system protocols to stabilize the extractable work of a charged battery through quantum control techniques \cite{santosStableAdiabaticQuantum2019,quachUsingDarkStates2020,gherardiniStabilizingOpenQuantum2020,mitchisonChargingQuantumBattery2021,rodriguezOptimalQuantumControl2022,YaoPRE2022}. 
\begin{figure*}
	\includegraphics[width=\textwidth]{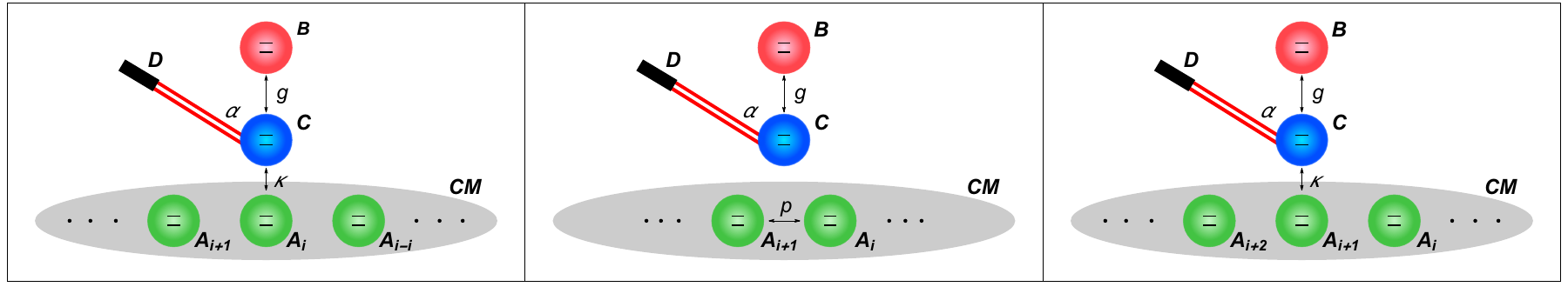}
	\label{f:schema}
	\caption{A collisional model of an OQB interacting with a non-Markovian environment: (left) two qubit systems representing a battery B (red qubit) and a charger C (blue qubit) interact between themselves for a discrete time $\delta t$ via a energy-exchange Hamiltonian, with coupling constant $g$. At the same time, C is driven by an external laser (with frequency $\alpha$) and interacts with the $i$-th environmental ancilla $A_i$, with coupling strength $\kappa$; (middle) before the next interaction between system and environment, the $i$-th and the $(i+1)$-th ancillas interact with each other via an incoherent partial-SWAP operation with probability $p$; (right) the interaction Hamiltonian is turned on for another time step $\delta t$ and the charger is put in contact with the next ancillary system.}

\end{figure*}

In most of the studies mentioned above, the dynamics induced by the environment is Markovian,
that is, the memory effects are negligible \cite{Breuer2002,Rivas2012}. 
This is indeed a useful assumption that simplifies considerably the characterization of the dynamics,
but in many circumstances one does need to go beyond it.
When the interaction of the system of interest with any further degree of freedom affecting its evolution
is not weak or  whenever the evolution of the environment takes
place on a similar time scale as the one of the system, one should take memory effects into account,
thus entering into the realm of non-Markovian quantum dynamics \cite{Rivas2014,Breuer2016,DeVega2017}.

As regards OQBs, memory effects have been considered only in very few cases. In \cite{carregaDissipativeDynamicsOpen2020} the authors study the behaviour in time of the energy stored in a qubit coherently driven and whose dissipative dynamics is described via specific non-Markovian master equations, without however discussing the effect on the ergotropy, that is on the maximum amount of actual work extractable from the battery qubit. In \cite{kaminNonMarkovianEffectsCharging2020} the dynamics of a system composed of a qubit-battery and a qubit-charger, with the second initially prepared in an excited state, is considered; there, both systems are interacting with two specific non-Markovian environments, where the initial excitation is eventually lost, and the analysis is focused on the time evolution of the ergotropy characterizing the battery. Finally, in \cite{ghosh_PRA2021} the charging and discharging of a qubit-battery is studied, and either a Markovian or non-Markovian environment plays the role of the charger. 

In this work we rather consider the following  model of an OQB, that has already been put forward in \cite{farinaChargermediatedEnergyTransfer2019} and that is pictured in Fig. \ref{f:schema}: two qubits, corresponding respectively to  battery and charger, interact via an energy exchange Hamiltonian. The energy is injected into the system via a driving Hamiltonian applied to the charger. The charger qubit is also coupled to an environment that will cause decoherence and dissipation for the battery, leading eventually to the steady-state of the dynamics. At variance with \cite{farinaChargermediatedEnergyTransfer2019}, where the interaction with a memoryless environment was considered, we will here study 
the impact of an environment that can induce memory effects on the dynamics of the system.

For this purpose, we exploit a useful framework to study non-Markovianity in the quantum setting: collisional models (CMs) \cite{CiccarelloPra2013} (see Refs. \cite{ciccarelloQuantumCollisionModels2021,Campbell2021_CM,cattaneo_brief_2022} for more details on these models and their connection with quantum thermodynamics and multipartite quantum dynamics). 
In these models both the environment and time are discretized: the environment is indeed represented by a discrete set of ancillas that interact with the systems sequentially at discrete times. We remark that CMs have already been employed as a tool for studying the behaviour of OQBs \cite{seahQuantumSpeedupCollisional2021,salviaQuantumAdvantageCharging2022,landiBatteryChargingCollision2021,mayoCollectiveEffectsQuantum2022,barraEfficiencyFluctuationsQuantum2022,shaghaghiMicromasersQuantumBatteries2022}. 
However in most of these works the stream of ancillas that constitutes the CM plays the role of the charger, except in \cite{landiBatteryChargingCollision2021} where it plays the role of the battery itself. In our model, as represented in Fig. \ref{f:schema}, we instead exploit CMs in order to describe the dissipative environment interacting with the charger qubit.

The structure of this work as follows: in Sec. \ref{Quantum Batteries}, we  recall the notion of quantum batteries and the figures of merit 
we use to assess them. In Sec. \ref{Collisional Model}, we  describe the non-Markovian CM that we exploit to characterize the environment for our model of dissipative quantum battery. In Sec. \ref{sec:model}, we fix the microscopic details of the model we  study. In Sec. \ref{sec:results}, we  present the results in the case of both a discrete-time and a continuous-time evolution, while in Sec.~\ref{sec:non-Markovianity} the precise connection between non-Markovianity and ergotropy is investigated. Sec.~\ref{sec:conclusions} concludes the paper with a final discussion and some outlooks of our work.
\section{Quantum Batteries}
\label{Quantum Batteries} 
A quantum battery can be described by a $d$-dimensional system with Hamiltonian:
\begin{equation}
    \hat{H}_0 = \sum_{n=1}^d \varepsilon_n  | \varepsilon_n \rangle \langle \varepsilon_n | \label{spectralH}
\end{equation}
with non-degenerate energy levels such that $\varepsilon_n < \varepsilon_{n+1}$. The charging and extraction of work can be modeled trough a time-dependent control parameter regulating the interaction that describes the process, giving a time-dependent Hamiltonian that reads $\hat{H}(t) = \hat{H}_0 + V(t)$. To evaluate the maximum energy that can be extracted from a given state of the system one considers a discharging process that starts at the time $t=0$ and finishes after a time $t=\tau$ when the battery is fully empty \cite{allahverdyanMaximalWorkExtraction2004}. 

The system is driven from the unitary operator generated by the full Hamiltonian and the average extracted work is given by
	\begin{equation}
		W(\hat{U}(\tau),\rho_0)=\hbox{Tr}[ \hat{H}_0\rho_0] - \hbox{Tr}[\hat{H}_0\hat{U}(\tau)\rho_0\hat{U}^\dagger(\tau)]
	\end{equation}
	where $\rho_0$ indicates the initial state of the system. The maximum amount of work that on average can be extracted from the system, a quantity called ergotropy, is then given by 
	\begin{equation}
	\mathcal{E}(\rho_0) = \max_{\hat{U}\in SU(d)} W(\hat{U}(\tau),\rho_0).
	\end{equation}
	
To rewrite it in a more operational form, we can consider the spectral decomposition of $\rho_0$, written as
	\begin{equation}
		\rho_0=\sum_{j=1}^d r_j | r_j \rangle \langle r_j |,
	\end{equation}
with $r_j \geq r_{j+1}$, and referring to Eq.~(\ref{spectralH}), it becomes evident that the state minimizing the system energy, that is, the equilibrium state once all the work has been extracted, is given by
	\begin{equation}
		\rho_0= \sum_{j=1}^d r_j | \varepsilon_j \rangle \langle \varepsilon_j |.
	\end{equation} 
	
As work can no longer be extracted from this state,  one usually refers to it as passive state. The ergotropy can simply be evaluated as the energy lost by the system during a discharging process that drives the system to its passive state:
	\begin{equation}
		\mathcal{E}(\rho_0)= \sum_{n,j=1}^d r_j \varepsilon_n (\left| \langle r_j \middle| \varepsilon_n \rangle \right|^2 - \delta_{jk}).
		\label{eq:ergotropy}
	\end{equation}

We remark that for a qubit state, and assuming the Hamiltonian $\hat{H}_0 = (\omega_0 / 2) \hat{\sigma}_z$, one can easily evaluate evaluate average energy and ergotropy in terms of the average values of the corresponding Pauli matrices, as
\begin{align}
    E(\rho) &= \frac{\omega_0}{2} \langle \hat{\sigma}_z \rangle \,,  \label{eq:E_qubit}\\
	\mathcal{E}(\rho) &= \frac{\omega_0}{2} (\left| \langle \hat{\boldsymbol\sigma} \rangle\right| + \langle \hat\sigma_z\rangle ) \,,  \label{eq:Erg_qubit}
\end{align}
where we have introduced the notation $\langle \hat O \rangle = \hbox{Tr}[\hat{O}\rho]$ and we have defined the quantity $\left| \langle\hat{\boldsymbol\sigma} \rangle\right|=\sqrt{\langle\hat\sigma_x\rangle^2+\langle\hat\sigma_y\rangle^2+\langle\hat\sigma_z\rangle^2}$. By observing the formula for the qubit ergotropy, and by recalling that the purity of a qubit state is equal to
\begin{align}
    \mu(\rho) = \Tr[\rho^2] = \frac{1+\left| \langle\hat{\boldsymbol\sigma} \rangle\right|^2}{2} \,,
\end{align}
we find that ergotropy can be expressed as a function of energy and purity as
\begin{align}
    \mathcal{E}(\rho) &= E(\rho) + \frac{\omega_0}{2}\sqrt{2 \mu(\rho) - 1}\,.
    \label{eq:ergoenergypurity}
\end{align}
It is thus clear that for qubits larger values of ergotropy can be obtained by maximizing not only the energy but also the purity of the state.
\section{Collisional Models}
\label{Collisional Model}
In this section we provide the basic notions on CMs, describing the differences between Markovian and non-Markovian scenarios, and discussing the continuous-time limits of such models.

\subsection{Discrete-time collisional models}\label{sec:dtc}

In a collisional model, a quantum system $S$ described by an Hamiltonian $\hat{H}_s$, is coupled to an environment $E$, made up from an infinite, but discrete, collection of subsystems, called ancillas, each with its own free Hamiltonian, such that the environment Hamiltonian reads $\hat{H}_E = \sum_i \hat{H}_{a_i}$, with $\hat{H}_{a_i}$ denoting the free Hamiltonian of each ancilla.
Time is also discretized, as the system interacts subsequently with each ancilla for a time interval $\delta t$, and one can thus introduce a collision rate $\gamma = 1/\delta t$. By denoting with $\hat{V}_{s,a_i}$ the interaction Hamiltonian between the system and the $i$-th ancilla, the unitary operator describing the evolution at the $i$-th step reads (we assume $\hbar=1$ throughout the manuscript):
\begin{equation}
	\hat{U}_i=e^{-i (\hat{H}_s + \hat{H}_{a_i} + \hat{V}_{s,a_i} ) \delta t }.
	\label{eq:unitary}
\end{equation}

A Markovian collisional model is realized when all these conditions are satisfied: i) there are no initial correlations between the system and the ancillae; ii) the initial state of the ancillae is factorized: $\hat{\sigma}_e = \bigotimes_i \eta_i$ (for simplicity we  also assume that they are prepared in the same initial state $\eta_i = \eta$); iii) there are no ancilla-ancilla collisions; iv) each ancilla collides only once with the system. 

Under these assumptions, the evolution of the state of the system can be described through a discrete map
\begin{equation}
	\rho_n=\mathcal{C}[\rho_ {n-1}] = \hbox{Tr}_{a_n}[\hat{U}_n(\rho_{n-1}\otimes\eta)\hat{U}_n^\dagger],\label{markovianmap}
\end{equation}

so that the state after $n$ steps is related to the initial state $\rho_0$ by
\begin{equation}\label{eq:semi}
    \rho_n=\mathcal{C}^n[\rho_ {0}].
\end{equation}
This corresponds, in the context of discrete evolutions, to the semigroup property,
which defines Markovian dynamics that are also homogeneous in time \cite{Breuer2002}.\\

To introduce non-Markovianity in the model, one has to relax at least one of the conditions listed above; here we introduce ancilla-ancilla collisions as shown in Fig. \ref{f:schema}, in the form of a ({\it incoherent}) partial-swap quantum map 
\begin{equation}
	\mathcal{W}_{n,n-1} = (1-p) \mathcal{I} +p\mathcal{S}_{n,n-1} \label{swap}
\end{equation} 
where $\mathcal{I}$ represents the identity map, and $p$ represents the probability of applying the SWAP operation $\mathcal{S}_{n,n-1}$ between the $(n-1)$-th and the $n$-th ancilla (as expected one has $0\leq p \leq 1$). In practice, before colliding with the system $S$, each ancilla with probability $p$ swaps its quantum state with the state of the previous ancilla that has just interacted with $S$. Clearly for $p=0$ one obtains the Markovian case previously discussed, while for $p>0$ one introduces memory effects in the environment.
 In particular, in the opposite limit of $p=1$ the system effectively interacts continuously with the same ancilla, and it is easy to show that its dynamics is described by the map
\begin{equation}
	\mathcal{F}_n[\rho_0]=\hbox{Tr}_{a_1}[\hat{U}_1^n (\rho_0 \otimes \eta) \hat{U}_1^{\dagger n}]. \label{nm_map}
\end{equation}
Through the maps $\mathcal{F}_j$ one can thus express the state of the system at the $n$-th step for arbitrary values of $p$ as \cite{ciccarelloQuantumCollisionModels2021}
\begin{equation}
	\rho_n= (1-p) \sum_{j=1}^{n-1} p^{j-1} \mathcal{F}_j [\rho_{n-j}] + p^{n-1} \mathcal{F}_n [\rho_0]. \label{nm_solution}
\end{equation}
\subsection{Continuous-time limit} \label{s:CMcontinuous}
We  now describe two examples where one can derive a continuous-time limit for the Markovian and non-Markovian collisional models we have just described. Deriving the continuous-time limit corresponds to taking the limit for the ancilla-system interaction time $\delta t \to 0$.  For simplicity and as it will correspond to the case we consider in the next sections, we assume that both the system and all the ancillas are qubits and we focus on a particular interaction Hamiltonian between system and ancillas.\\

We start from the Markovian case and we consider as a paradigmatic example the following system-ancilla interaction Hamiltonian
\begin{align}
\hat{V}_{s,a_i} = \sqrt{\frac{\kappa}{\delta t}} \left(\hat{\sigma}_+^{ (s)} \hat{\sigma}_-^{ (a_i)}  + \hat{\sigma}_-^{ (s)} \hat{\sigma}_+^{ (a_i)} \right) , \label{eq:ancillainteraction}
\end{align}
with $\kappa\geq0$ and where $\hat{\sigma}_- = (\hat\sigma_x + i \hat\sigma_y)/2$, and $\hat{\sigma}_+=(\hat{\sigma}_-)^\dag$.
The introduction of a coupling constant $\xi = \sqrt{\kappa/\delta t}$ that diverges in the limit $\delta t \to 0$  is necessary to obtain a well-defined continuous limit \cite{ciccarelloQuantumCollisionModels2021,gross_qubit_2018}. In fact, 
by considering all the ancillas in the initial state $\eta = |0\rangle_i {}_i\langle 0|$ (such that $\hat{\sigma}_-^{ (a_i)}|0\rangle_i = 0$), 
the dynamics described by the collisional model in the limit $\delta t \to 0$ is equivalent to the one described by the Markovian master equation in the Lindblad form \cite{lindbladGeneratorsQuantumDynamical1976,Gorini:1975nb}
\begin{equation}\label{eq:lindblad}
\frac{d\rho}{dt} =  - i [ \hat{H}_s, \rho] + \kappa \mathcal{D}[\hat{\sigma}_-]\rho  \,,
\end{equation} 
where we have defined the superoperator $\mathcal{D}[\hat{c}]\rho = \hat{c} \rho \hat{c}^\dag - (\hat{c}^\dag \hat{c} \rho + \rho \hat{c}^\dag \hat{c})/2$. \\

The continuous-time limit of the non-Markovian collisional model introduced above is not as straightforward. In fact it is possible to numerically check that, by introducing the partial-swap quantum map defined in Eq. (\ref{swap}) with a fixed value of $p$, and by decreasing the time-step interval $\delta t$, the dynamics does not converge to a well-defined continuous-time dynamics. 

One can however show that a continuous time limit can be taken by introducing a memory rate $\Gamma$, and by making the partial-swap parameter $p$ time-dependent as $p=\exp{-\Gamma \delta t}$ \cite{ciccarelloQuantumCollisionModels2021}. In particular, in the limit for $\Gamma \delta t \ll 1$, such that $p \approx 1-\Gamma \delta t$, a memory-kernel non-Markovian integro-differential ME can be found in the form
\begin{equation}
    \label{eq:integroME}
	\dot{\rho}=\Gamma \int_0^t dt' e^{-\Gamma t'}\mathcal{F}(t')[\dot{\rho}(t-t')] + e^{-\Gamma t} \dot{\mathcal{F}}(t)[\rho_0] \,,
\end{equation}
where $\mathcal{F}(t)$ is the continuous-time version of the map (\ref{nm_map}).
\section{A Markovian open quantum battery}\label{sec:mm}
\label{sec:model}
In the following, we  use as benchmark of our investigation the qubit model of OQB in a Markovian environment studied in \cite{farinaChargermediatedEnergyTransfer2019}, which thus fixes the microscopic details of the quantum battery under study.
The qubit $B$ represents the quantum battery itself, with free Hamiltonian $\hat{H}_{B,0} = (\omega_0 / 2) \hat{\sigma}_z^{ (B)}$, and it interacts with another qubit $C$ that corresponds to the charger, via an energy-exchange interaction
\begin{align}
	\hat{H}_{BC}= g \left(\hat{\sigma}_-^{ (B)}\hat{\sigma}_+^{(C)} + \hat{\sigma}_+^{(B)}\hat{\sigma}-^{(C)}\right). 
\end{align}
The Hamiltonian for the charger is the sum of two terms, $\hat{H}_C= \hat{H}_{C,0} + \hat{H}_{\sf drive}$, where $\hat{H}_{C,0} = (\omega_0 / 2) \hat{\sigma}_z^{(C)}$ is the free Hamiltonian of the charger, while
\begin{align}
\hat{H}_{\sf drive} = 
\alpha \left(e^{-i\omega_0 t}\hat{\sigma}_+^{(C)}  + e^{+i\omega_0 t}\hat{\sigma}_-^{(C)} \right)
\end{align}
represents the driving of the charger qubit, injecting energy in the system via a laser driving at frequency $\omega_0$.

In \cite{farinaChargermediatedEnergyTransfer2019}, 
the charger is subjected to an amplitude damping due to the interaction with a 
memoryless environment. By going to the interaction picture with respect to the Hamiltonian $\hat{H}_0 = \hat{H}_{B,0} + \hat{H}_{C,0}$, the dynamics of the battery and charge state is described by the Lindblad equation
\begin{align}
\frac{d\rho}{dt} = - i [ \hat{H}^\prime_{BC}, \rho] + \kappa \mathcal{D}[\hat{\sigma}_-^{(C)}]\rho \,, \label{eq:QBME}
\end{align}
where 
\begin{align}
\hat{H}^\prime_{BC}  =g\left(\hat{\sigma}_-^{ (B)}\hat{\sigma}_+^{(C)} + \hat{\sigma}_+^{(B)}\hat{\sigma}-^{(C)}\right) + \alpha\,\hat{\sigma}_x^{(C)} \,. \label{eq:hbcp}
\end{align}
The global maximum of the ergotropy is reached in the large-loss limit (i.e. for $\kappa \gg \alpha $), and by tuning driving and coupling such that $\alpha \approx 1.09 g$, yielding \cite{farinaChargermediatedEnergyTransfer2019}  
\begin{align}
    \mathcal{E}_{\tiny{\sf max }} = \frac{\sqrt{2}-1}{2}\omega_0 \approx 0.207 \omega_0 \,.
    \label{eq:maxergo}
\end{align}
\\

The dynamics described by the master equation (\ref{eq:QBME}) corresponds to the continuous-time limit of a particular Markovian collision-based model, as  follows from what shown in Sec. \ref{s:CMcontinuous}. Here, we explore what happens beyond this regime by 
taking into account memory effects induced by the environment, first
via a discrete-time collisional model, and then 
looking at the continuous-time limit.
As in \cite{farinaChargermediatedEnergyTransfer2019} we  focus on the steady-state properties of the battery $\rho^{(B)}_{ss}$, evaluating via Eqs.~(\ref{eq:E_qubit}) and (\ref{eq:Erg_qubit}) the corresponding average energy $E_{ss}$ and ergotropy $\mathcal{E}_{ss}$.
\section{Results}
\label{sec:results}
\subsection{Discrete-time collisions}
We first study the discrete-time collisional model where the open system consists of the two qubits representing the battery and
the charger, with Hamiltonian $\hat{H}_S = \hat{H}^\prime_{BC}$ as in Eq.(\ref{eq:hbcp}),
and the charger interacts
with the ancillas describing the environment, via a Hamiltonian of the form (\ref{eq:ancillainteraction}), which in our case reads
\begin{align}
\hat{V}_{s,a_i} = \sqrt{\frac{\kappa}{\delta t}} \left(\hat{\sigma}_+^{(C)} \hat{\sigma}_-^{ (a_i)}  + \hat{\sigma}_-^{(C)} \hat{\sigma}_+^{ (a_i)} \right).
\end{align}
Each system-ancilla collision lasts for a time $\delta t = 1/\gamma$, leading to a unitary evolution as in Eq. (\ref{eq:unitary}), and we introduce ancilla-ancilla interactions via the partial-swap map defined in Eqs. (\ref{swap}). 

We now present our results for the steady state of the battery, focusing on the role of the parameter controlling the information backflow 
to the battery and the charger, that is the swap-probability $p$, as well as on 
the battery-charger coupling $g$ and the driving constant $\alpha$. 
We have considered collision rates $\gamma$ much larger than the other frequencies characterizing the dynamics, so that by setting the swap probability $p=0$ 
we (numerically) recover the continuous-time limit dynamics described by Eq. (\ref{eq:QBME}).
\begin{figure*}
	\centerline{\includegraphics[width=1.05\textwidth]{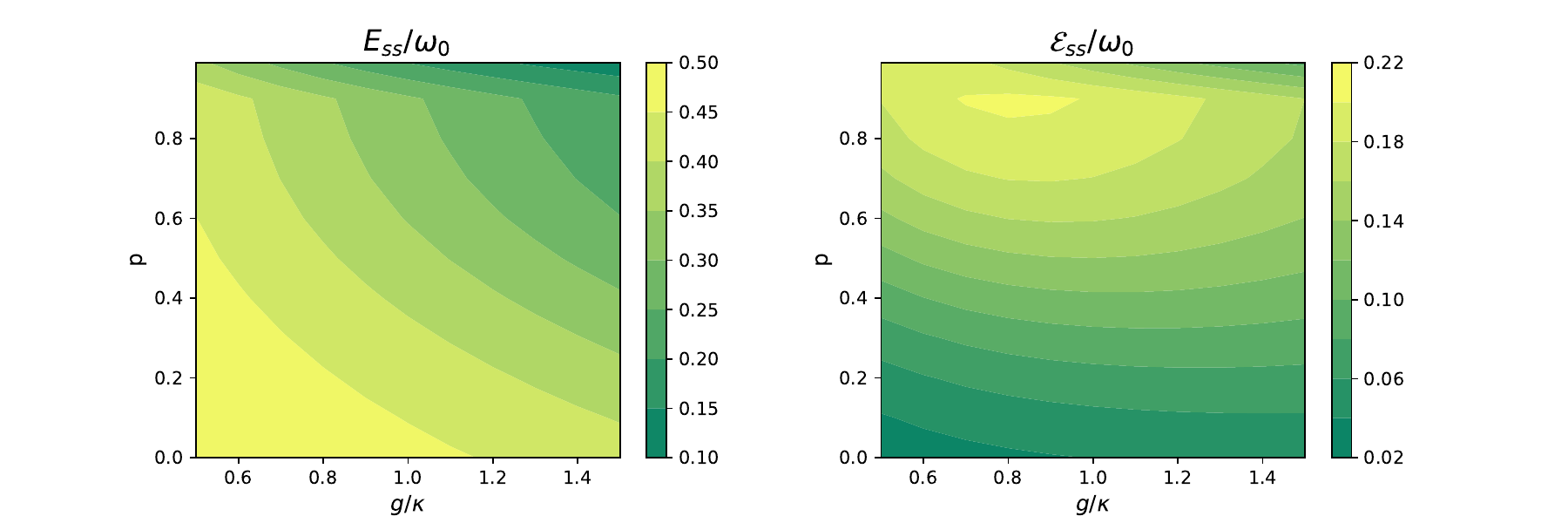}}
	\caption{Steady-state value for the battery energy $E_{ss}/\omega_0$ (left) and ergotropy $\mathcal{E}_{ss}/\omega_0$ (right) as functions of $g/\kappa$ and $p$; with $\alpha=0.9\kappa$ and $\gamma= 10^2 \kappa$. While the energy is monotonously decreasing for both $p$ and $g$ increasing, the ergotropy displays a more complex behaviour. As a function of $g$, it has a maximum value for $\alpha \approx 1.09 g$, which, for our choice of parameters, can be seen in the central area of the plot. As a function of $p$, it has a non-monotonous behaviour, increasing until it reaches a maximum value for $p \approx 0.9$ before starting to decrease.}
		\label{f:EWRcontour2}
\end{figure*}

To characterize the battery charging properties in an environment with memory and beyond the large-loss limit, in Fig. \ref{f:EWRcontour2} we plot the steady-state ergotropy and energy as a function of both $p$ and $g$, by fixing the value of $\alpha=0.9\kappa$. 
We immediately notice a different behaviour between energy and ergotropy: in particular, while the former shows a generally decreasing behaviour with $p$, the ergotropy shows a non-monotonous behaviour as a function of both $p$ and $g$. However, a noticeable optimal region of parameters can be identified, corresponding to $p\approx 0.9$ and $g\approx \alpha$, where the maximum amount of ergotropy is observed. This different behaviour emphasizes the necessity to use a measure of extractable work like the ergotropy as figure of merit for a quantum battery instead of simply evaluating its energy. Indeed, for fixed values of $\alpha/\kappa$ and $g/\kappa$, increasing $p$ can lead to an increase in the portion of the system maximum extractable energy, its ergotropy, while decreasing the maximum amount of average energy itself. As we remarked in Eq. (\ref{eq:ergoenergypurity}), the ergotropy of a qubit can be expressed as a function of the energy and of the purity of the state. The previous observations clearly hint to the fact that the ergotropy enhancement due to the backflow of information from the environment corresponds to the generation of less mixed steady-states.

\begin{figure}
    	\includegraphics[width=0.5\textwidth]{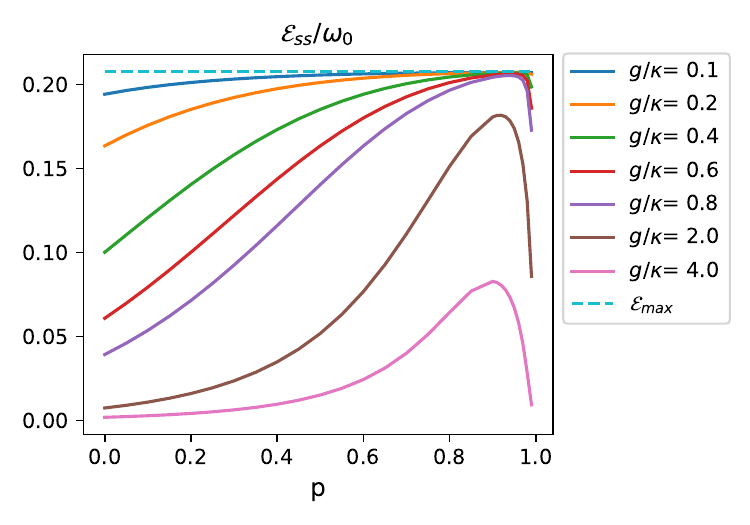}
	\caption{ Steady-state ergotropy $\mathcal{E}_{ss}/\omega_0$ as function of $p$ for different values of $g/\kappa$ and fixing the coupling parameter such that $\alpha=1.09 g$ and  $\gamma/\kappa=10^2$. The value for the steady-state ergotropy is upper bounded by the maximum found in the memoryless case in the large-loss limit \cite{farinaChargermediatedEnergyTransfer2019}. Indeed, for $p=0$, the ergotropy is the higher the lower is $g/\kappa$, but when p increases, the ergotropy also starts to increase for all curves at different speed, with the curves further from the large-loss limit exhibiting the largest increment, all plateauing under the boundary condition. Therefore systems that are already in the optimal region of parameters display a mostly flat behaviour, while systems out of that region can achieve the maximum gain from a backflow of information due to the environment.    }
	\label{f:ergotropyfixedalpha}
\end{figure}

To further understand this improvement of the ergotropy due to memory effects, 
we compare the maximum of ergotropy in the region with $p>0$ with
the maximum achieved in the large-loss limit at $p=0$.
In Fig. \ref{f:ergotropyfixedalpha} we therefore show the behaviour of the steady-state ergotropy as a function of the memory parameter $p$, for different values of the coupling $g$, and by fixing the driving parameter $\alpha$ such that the optimal condition $\alpha=1.09g$, identified for the large loss memoryless scenario \cite{farinaChargermediatedEnergyTransfer2019}, is satisfied. We remark that in general, at each value of $p$, a different optimal tuning condition between $\alpha/\kappa$ and $g/\kappa$ can be found. This tuning condition, as we will later describe, is generally close to the value at $p=0$, and also the difference in the corresponding values of the ergotropy is negligible, as we have numerically verified for all the parameter regimes considered in our plots. 
\begin{figure*}
	\centerline{\includegraphics[width=\textwidth]{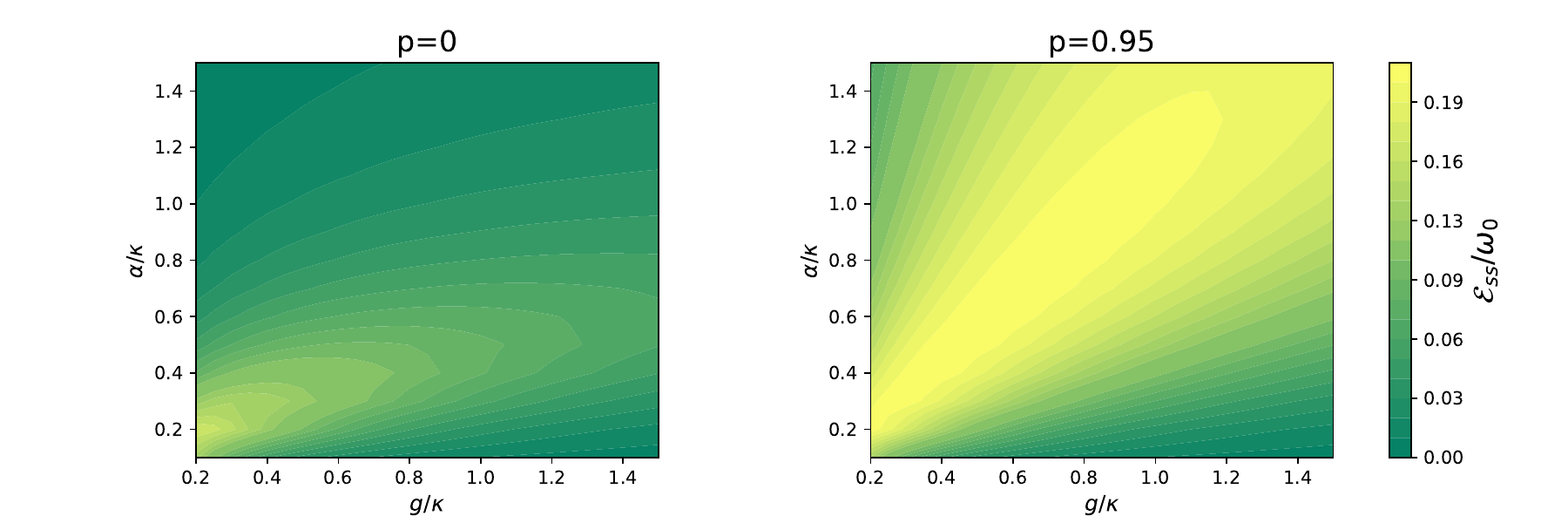}}
	\caption{Steady-state ergotropy $\mathcal{E}_{ss}$ as function of $\alpha$ and $g$ for fixed values of $p$ (left: $p=0$; right: $p = 0.95$); $\gamma=10^{2}\kappa$. On the left, the steady-state ergotropy shows the same result obtained in \cite{farinaChargermediatedEnergyTransfer2019}, i.e., the ideal tuning condition to maximize the ergotropy exists in the large-loss limit, $\alpha \ll \kappa$, for a fixed ratio of $\alpha$ and $g$. On the right, the same plot realized for a higher value of $p$, shows how a backflow of information from the environment can extend the optimal tuning region also outside the large-loss limit.}
		\label{tuning}
\end{figure*}

We observe that for small values of $\alpha$ and $g$, that is towards the large-loss limit, the behaviour of the ergotropy as a function of $p$ is almost flat. Only for larger values of $g$ and $\alpha$ we observe a more evident non-monotonous behaviour of the ergotropy as a function of $p$, and that an enhancement is observed with respect to the Markovian case $p=0$, reaching a maximum for a certain, relatively large, value of $p$. 

The most remarkable result we observe in Fig.~\ref{f:ergotropyfixedalpha} is that 
the maximum amount of ergotropy $\mathcal{E}_{\tiny {\sf max}}$ derived in the memoryless case in the large-loss regime (see Eq. (\ref{eq:maxergo})) maximizes the ergotropy also in our collisional model with memory, but $\mathcal{E}_{\tiny {\sf max}}$ can be achieved also beyond 
the large-loss regime in the presence of large enough values of $p$. As we numerically checked, the maximum value
of ergotropy is indeed reached in all the considered regimes via the same steady state,
described by Bloch vector components
 $\langle \hat{\sigma}_x\rangle=-\sqrt{\sqrt{2}-1}$, $\langle \hat{\sigma}_y\rangle=0$, $\langle \hat{\sigma}_z\rangle=\frac{1}{\sqrt{2}}-1$.
Importantly, for any value of $g/\kappa \lesssim 1$, and for the right tuning condition between $g$ and $\alpha$, there is always a value of $p$ for which the ergotropy approximates its maximum value, while this is no longer the case for larger values of $g/\kappa$. 

We also observe that in our model only coherence contributes to ergotropy, and as a consequence its maximum possible value a priori could be $\omega_0/2$~\cite{francicaQuantumCoherenceErgotropy2020}. However, we find $\mathcal{E}_{\tiny {\sf max}}<\omega_0/2$ irrespectively of the values of the parameters considered and of the memory properties of the environment, hinting to the fact that the optimal performance of the battery depends on the sole operatorial form of the interaction between battery and charger and between charger and environment.

We then further investigate the optimal tuning condition between $\alpha$ and $g$ to achieve maximum ergotropy. In Fig. \ref{tuning} we plot the steady-state ergotropy as a function of $\alpha$ and $g$, respectively, in the model without swap, $p=0$, and for a large
swap probability $p=0.95$. In the former case, we observe as expected that the maximum of the ergotropy is obtained in the large-loss limit $\alpha \ll \kappa$ and for $\alpha=1.09 g$, as analytically demonstrated in \cite{farinaChargermediatedEnergyTransfer2019}.
On the other hand, in the case
of a collisional model with memory, we now observe an extended region of parameters where one can find large values of the ergotropy, and in particular there exist a linear boundary between the parameters $g/\kappa$ and $\alpha/\kappa$ along which the maximum value of the ergotropy can be found for a limited region of the plot, approximately identified by the condition $g/\kappa \lesssim 1$. We can thus conclude that the presence of memory effects
allows us to obtain the maximum value of ergotropy $\mathcal{E}_{\tiny{\sf max}}$ for a larger region in the parameters space.

\begin{figure}
    	\centerline{\includegraphics[width=0.5\textwidth]{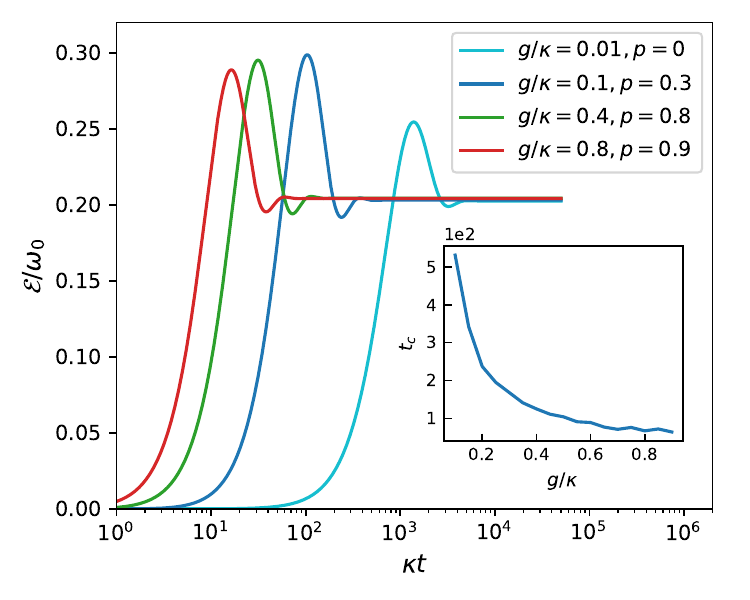}}
	\caption{Ergotropy as a function of time for different combinations of parameters. For all curves with given values of $g/\kappa$ and $\alpha/\kappa$, the swap probability $p$ has been chosen as the minimum needed to approximate $\mathcal{E}_{\tiny {\sf max}}$ within one percent of relative error. The inset shows the charging time $t_c$ as a function of $g/\kappa$, where the other parameters are chosen following the same logic of the main plot (the charging time $t_c$ is defined as the time necessary for the system to reach its steady state).}
	\label{f:dynamics}
\end{figure}

Besides the steady-state properties of the ergotropy of the battery, also its transient evolution can be indeed of interest,
for example because the maximum value can be obtained on different timescales depending on the parameters fixing the dynamics.
In Fig.~\ref{f:dynamics}, we plot the ergotropy as a function of time for both the memoryless model in the large loss limit ($g/\kappa=0.01$) and various combinations of the values for the swap probability $p$ and for $g/\kappa$; in all these cases $\alpha/\kappa$ is determined by the optimal tuning condition that maximizes the steady state ergotropy. What is shown is that the various curves reach approximately
the same asymptotic value $\mathcal{E}_{\tiny{\sf max }}$, but this happens the sooner the larger $g/\kappa$. 
This shows that, if larger values of the coupling $g$ are possible, an accurate choice of $p$ can be done in order to speed up the dynamics and obtain the maximum amount of the ergotropy in a shorter time.
Indeed, one should keep in mind that too large values of $p$ will eventually slow down the stabilization process, as in the limit $p\to 1$ one obtains a unitary dynamics. However a large enough value can still be chosen to approximate the maximum ergotropy at steady state and still speed up the dynamics. For this reason the values of $p$ leading to the curves in Fig.~
\ref{f:dynamics} correspond to the minimum value needed, at fixed $g/\kappa$, to approximate the maximum state ergotropy. In the inset we plot the time $t_c$ that is required for these curves to reach their steady-state value as a function of g: we indeed observe how larger values of g monotonously lead to a faster charging time.\\

\subsection{Continuous-time limit dynamics}
As we described in Sec. \ref{s:CMcontinuous}, the considered collision-based model with memory leads to a well defined continuous-time limit if we introduce a memory rate $\Gamma$, allow the partial-swap probability to be dependent on the collision rate as $p=e^{-\Gamma/ \gamma}$, and consider the regime $\Gamma\ll \gamma$. In fact under these conditions, the limit $1/\gamma = \delta t \to 0$ of the collision-based model is well described by the integro-differential ME (\ref{eq:integroME}), where the map $\mathcal{F}(t)$ also includes the effect of the driving Hamiltonian acting on the charger qubit and the charger-battery coupling Hamiltonian. 

\begin{figure*}
	\centerline{\includegraphics[width=1.\textwidth]{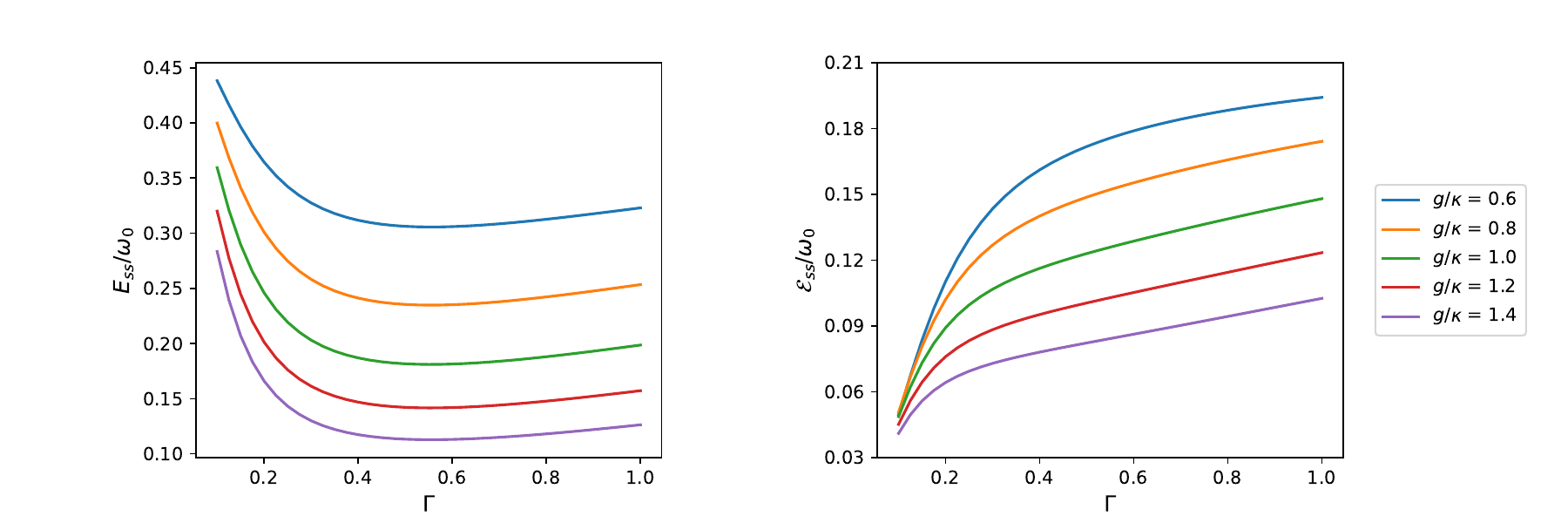}}
	\caption{Steady-state energy $E_{ss}/\omega_0$ (left) and ergotropy $\mathcal{E}_{ss}/\omega_0$ (right) for the continuous-time limit of the collision based model described by the ME (\ref{eq:integroME}) as a function of the memory rate $\Gamma$. Different curves correspond to different values of the coupling $g$. Other parameters are fixed as $\alpha=0.9\kappa$ and $\gamma=10^3 \kappa$ (equivalent results are obtained as expected for larger values of $\gamma$, that is towards the continuous-time limit).
    As functions of the memory rate $\Gamma$, energy and ergotropy show opposite behaviours, with the energy decreasing to a minimum as $\Gamma$ increases, while the ergotropy monotonously increases. In general, for higher values of $\Gamma$, more energy passed onto the system can no longer be recovered. This means, in this regime of parameters, that the backflow of information from the environment has a negative impact on the properties of the battery.}
	\label{f:EWRcontinuous}
\end{figure*}

We have thus exploited our collisional model to simulate such 
integro-differential master equation, by considering the regime $\Gamma \ll \gamma$ and by fixing $\gamma/\kappa = 10^3$  (we have checked numerically that equivalent results are obtained by considering larger values of $\gamma$ and thus approaching the continuous-time limit $\delta t \to 0$). The results in terms of energy and ergotropy for the steady state of the dynamics are reported in Fig. \ref{f:EWRcontinuous}. In particular, when compared to the discrete-time model where the energy decreases when increasing $p$, we observe that the energy has a minimum as a function of the memory rate $\Gamma$. On the other hand the ergotropy, which is our main figure of merit, monotonically increases with $\Gamma$.
This result may seem in contradiction with what we have observed in the previous section for the discrete-time model, as it hints that ergotropy increases when one moves towards the Markovian regime, that is by increasing $\Gamma$. However we have to remind ourselves that in order to obtain a well-defined continuous limit, we are restricting to values of $p$ very close to one; consequently, even if a one-to-one correspondence between the discrete- and continuous-time models cannot be properly defined due to the dependence between $p$, $\Gamma$ and $\gamma$, what we observe in Fig. \ref{f:EWRcontinuous} approximately corresponds to the region in Fig. \ref{f:EWRcontour2} where the ergotropy, after having reached its maximum, decreases when $p$ takes large values and approaches one.

Until now, we have studied the steady state ergotropy for different values of the parameter $p$, which quantifies 
the probability of having collisions among the environmental ancillas, in turn inducing memory effects
in the evolution of the battery and the charger as discussed in Sec.~\ref{sec:dtc}.
On the other hand, to fully understand the role of non-Markovianity in the properties of the battery
we need to quantify in an explicit way the amount of memory effects in the dynamics, i.e., to introduce
a measure of non-Markovianity. This is what we are going to do in the next section.

\section{The role of non-Markovianity}
\label{sec:non-Markovianity}

Among the different quantifiers of non-Markovianity \cite{Rivas2014,Breuer2016}, we exploit the so-called geometrical measure of non-Markovianity, defined in \cite{Lorenzo2013}. The amount of non-Markovianity in a given dynamics is here quantified by integrating the expansions of the volume of possible states accessible by the system throughout the evolution. 
Compared to other measures of non-Markovianity -- such as those based 
on non-monotonic behaviors of the trace distance \cite{Breuer2002}
or on the breaking of divisibility \cite{Rivas2010} --
the geometrical measure is weaker, i.e., it might be equal to zero also for dynamics where other measures are not.
On the other hand, besides its clear geometrical meaning that will be recalled below, it has the advantage of being manageable also for high-dimensional systems: in fact, it provides us with a clear computational convenience already for the two-qubit system formed by the battery and the charger.
\subsection{Geometrical measure of non-Markovianity}
Every state $\rho$ of a finite $n$-dimensional system, i.e., positive trace-one linear operator on $\mathbbm{C}^n$,
can be written as \cite{Kimura2003}
\begin{equation}\label{eq:rho}
    \rho = \frac{1}{n}\left(\mathbbm{1}+\sum_{\alpha=1}^{n^2-1} r_\alpha \hat{G}_\alpha\right),
\end{equation}
where $\mathbbm{1}$ is the identity map on $\mathbbm{C}^n$, $r_\alpha \in \mathbbm{R}$
and the $\hat{G}_\alpha$ are the hermitian traceless generators of $SU(n)$.
For $n=2$, the latter identify with the Pauli matrices and Eq. \eqref{eq:rho} corresponds
to the well-known Bloch-vector representation of the qubit states \cite{Nielsen2000};
we then call the vector $\bm{r}$ with components $\left\{r_\alpha\right\}_{\alpha=1,\ldots, n^2-1}$
generalized Bloch vector.

Now, the map $\Lambda(t)$ describing the state evolution,
\begin{equation}
    \rho(t) = \Lambda(t)[\rho(0)],
\end{equation}
corresponds to an affine transformation of the generalized Bloch vector
\begin{equation}\label{eq:aff}
	\bm{r}(t)=A(t)\bm{r}(0) + \bm{q}(t),
\end{equation}
where $A(t)$ is a $n^2-1 \times n^2-1$ real matrix with elements
\begin{equation}
    A_{\alpha,\beta}(t)=Tr[\hat{G}_\alpha\Lambda(t)[\hat{G}_\beta]]
\end{equation}
and $\bm{q}(t)$ is a $n^2-1$ real vector.
The matrix $A(t)$ describes rotations (possibly composed with inversions)
and contractions of the generalized Bloch vectors, so that its determinant
$||A(t)||$ accounts for the contraction factor of the volume of accessible states, i.e., 
the volume $V(t)$ of the image of the set of generalized Bloch vectors under the action of the affine transformation defined in Eq. (\ref{eq:aff}).

The geometrical measure of non-Markovianity defined in \cite{Lorenzo2013} is given in terms of the variations of the volume
of the set of accessible states as
\begin{equation}
	\mathcal{N}_b = \frac{1}{V(0)}\int_{\frac{dV}{dt}>0} \frac{dV(t)}{dt} dt = \int_{\frac{d||A||}{dt}>0} \frac{d||A(t)||}{dt} dt,
\end{equation}
that is, as the integral of the derivative of the volume of accessible states over all the time intervals where such volume increases, divided by the overall volume of the set of states.
The basic idea is that in Markovian evolutions, such as semigroup dynamics, there is a monotonic decay of the volume of accessible states corresponding to an irreversible loss of information from the open system \footnote{In \cite{Lorenzo2013} this idea is expressed quantitatively in terms of the evolution of the entropy associated with a random distribution of the initial states.}, while the revivals in time of the volume of accessible states can be read as memory effects leading to a (partial) recovery of previously lost information.

\subsection{Memory effects and ergotropy}
\label{sub:memory_and_ergotropy}
\begin{figure}
	\includegraphics[width=0.5\textwidth]{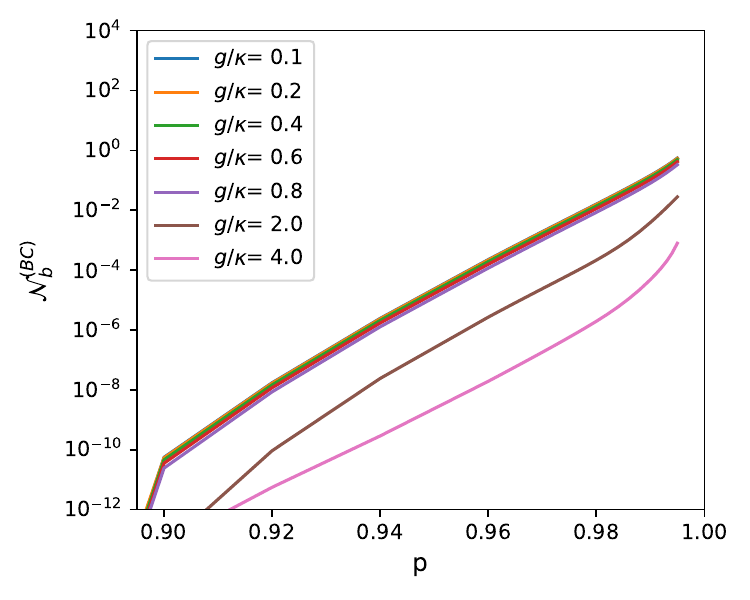}
	\caption{Geometrical measure $\mathcal{N}_b^{(BC)}$ for the composite system (charger + battery) as a function of $p$ for different values of $g$. The other parameters are $\alpha=1.09g$, $\gamma=10^2\kappa$.
    As expected, the geometrical measure grows monotonously with p in the area shown. We remark that for smaller values of $p$, the measure of non-Markovianity is zero. }
	\label{measure2}
\end{figure}

In the following we study the behaviour of the geometric non-Markovianity measure  $\mathcal{N}_b$, by focusing on the discrete-time scenario. In particular, we evaluate the measure for both the composite system (battery plus charger) $\mathcal{N}_b^{(BC)}$ and for the battery subsystem only $\mathcal{N}_b^{(B)}$. 

The non-Markovianity of the composite system, plotted in Fig. \ref{measure2}, follows the expected behaviour as it is monotonically increasing with the swap probability $p$, which in fact quantifies the memory of the environment affecting the composite system. We also observe that $\mathcal{N}_b^{(BC)}$ is equal to zero in a large region of values of $p$, the extension of which depends on the the different parameters involved in the dynamics.
\begin{figure}
 	\includegraphics[width=0.5\textwidth]{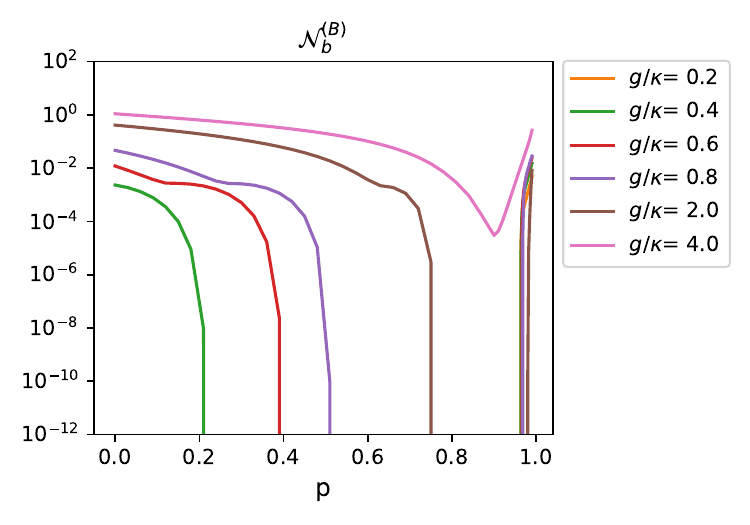}
 	\caption{Geometrical measure for the battery subsystem $\mathcal{N}_b^{(B)}$ as a function of $p$ for different values of $g$. The other parameters are: $\alpha=1.09g$, $\gamma=10^2\kappa$.
    At variance with the measure for the composite system, here $\mathcal{N}_b^{(B)}$ has a non-zero value for $p=0$, due to the memory effects caused by the battery-charger coherent interaction, but as the backflow of information from the environment to the charger increases, the measure itself decreases. For values of the ratio $g/\kappa$ small enough, $\mathcal{N}_b^{(B)}$ goes to zero for a given value of $p$, and it starts to increase only for very large values of $p$. For larger values of the ratio $g/\kappa$, $\mathcal{N}_b^{(B)}$ decreases to a finite minimum, without reaching zero. We remark that our numerical simulations show that towards the large-loss limit ($g/\kappa=\{0.1,0.2\}$), $\mathcal{N}_b^{(B)}$ is zero except in the region of very large values of $p$.}
 	 \label{measurebattery}
\end{figure}

If we rather focus on the battery subsystem only, the behaviour of the non-Markovianity measure $\mathcal{N}_b^{(B)}$ is rather different, as one can observe in Fig. \ref{measurebattery}. 
First, a non-zero value of non-Markovianity is obtained for a memoryless environment, that is for $p=0$.
A Markovian dynamics of the joint system consisting of the battery and the charger, as fixed by the semigroup map in Eq. (\ref{eq:semi}) (or the Lindblad equation (\ref{eq:lindblad}) in the continuous case), can well induce a non-Markovian evolution of the battery only: the two-fold exchange of information between the battery and the charger qubits due to their coherent interaction will generally result in a repeated backflow of information to the battery, with large values of non-Markovianity. Increasing the value of $p$
the non-Markovianity of the battery has a non-monotonic behavior, which strongly depends on
the ratio $g/\kappa$. For low values of $g/\kappa$ ($g \lesssim 2\kappa$), 
$\mathcal{N}_b^{(B)}$ decreases and reaches zero at different values of the swap probability $p$, and then suddenly increases for $p\gtrsim 0.95$. For higher values of the ratio $g/\kappa$, $\mathcal{N}_b^{(B)}$ still decreases, but it no longer reaches zero.

An intuitive interpretation of this behaviour 
traces back to the fact that for values of $p$ larger than zero the battery dynamics has two sources of memory. 
Besides the one already mentioned above originating directly in the coherent interaction with the charger, now there is also an information backflow from the environment, due to the ancilla-ancilla collisions.
As follows from the discussion in Sec.~\ref{sec:dtc}, the latter will store and give back memory about previous stages of the evolution, affecting in the first instance the charger, with which the ancillas directly interact, and then also the battery due to its interaction with the charger.
The overall behaviour of the battery dynamics depends on the interplay between these two sources of memory, so that the presence of a second source can reduce the impact of the memory effects induced by the charger observed in the case $p=0$. 
This effect strongly depends on the ratio of the two couplings $g$ and $\kappa$, connecting respectively the charger with the battery, and the environment with the charger. For small values of this ratio, the environment is capable of affecting the battery dynamics enough to cancel out the information backflow from the charger, while for higher values of the ratio the impact of the memory from the environment on the non-Markovianity of the battery is weaker.

Note that there are values of $p$ such that the joint battery-plus-charger dynamics is non-Markovian, 
while the reduced battery dynamics is Markovian, see Figs. \ref{measure2} and \ref{measurebattery};
this is indeed in contrast with what happens commonly, that is, by enlarging the set of degrees of freedom one moves
from non-Markovian to Markovian evolutions. On the other hand, a similar phenomenon has been observed in \cite{Laine2012} and
exploited for quantum teleportation in \cite{Laine2014}, where non-local memory effects in the dephasing dynamics
of two qubits -- originating from the presence of initial correlations within the environment -- affect the joint systen
made by the two qubits, but neither of the two individually.

By comparing Fig. \ref{f:ergotropyfixedalpha} and Fig. \ref{measurebattery}, we notice some 
significant correlations between $\mathcal{N}_b^{(B)}$ and $\mathcal{E}_{ss}$. 
The two have in general opposite behaviour, with one increasing while the other decreases. 
Even more, every time the steady state ergotropy approximate its absolute maximum $\mathcal{E}_{ss}$, the battery non-Markovianity $\mathcal{N}_b^{(B)}$ is equal to zero. For small values of the ratio $g/\kappa$ (e.g. $g/\kappa=\{0.1,0.2\}$), the steady state ergotropy is approximately equal to $\mathcal{E}_{ss}$ for almost any value of $p$, and similarly $\mathcal{N}_b^{(B)}$ is zero for most values of $p$. Analogously, the range of values for which the curves referred to $g/\kappa=\{0.4,0.6,0.8\}$ approximate the maximum of steady state ergotropy 
corresponds to a region without memory effects in the battery dynamics.
We also notice how, differently, the curve with $g/\kappa=2$ does not reach $\mathcal{E}_{ss}$ even though its value of $\mathcal{N}_b^{(B)}$ reaches zero for certain values of $p$. 

\begin{figure}
 	\includegraphics[width=0.5\textwidth]{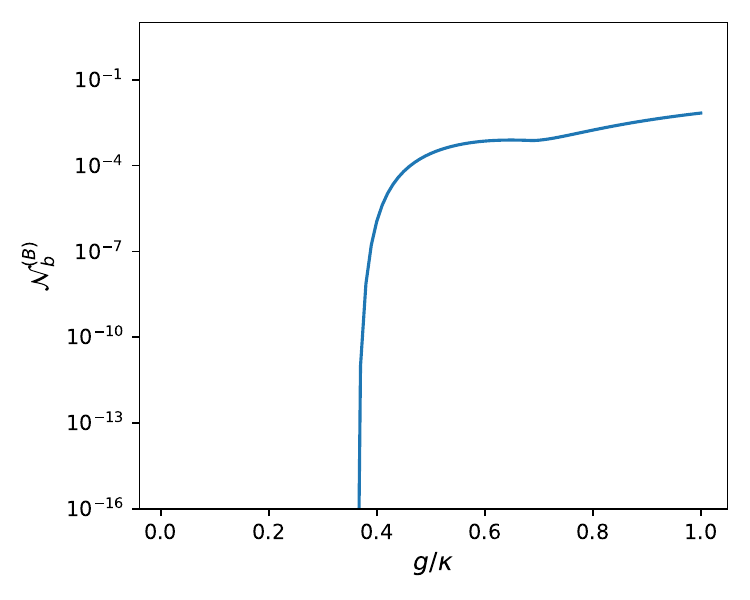}
 	\caption{Geometrical measure for the battery subsystem $\mathcal{N}_b^{(B)}$ as a function of $g/\kappa$ in the presence of a memoryless environment as discussed in \cite{farinaChargermediatedEnergyTransfer2019}. For each value of $g/\kappa$, the driving $\alpha$ has been chosen in order to optimize the steady-state ergotropy.}
 	 \label{f:Farina_nonMarkov}
\end{figure}

We conclude that the absence of memory effects for the battery dynamics seems to be necessary, although not sufficient, to reach the maximum of steady state ergotropy in our model. 
This is confirmed by the behaviour of $\mathcal{N}_b^{(B)}$ as a function of the ratio $g/\kappa$ for the scenario described by the Lindblad master equation considered in \cite{farinaChargermediatedEnergyTransfer2019} and that corresponds to our collisional model with no swap between ancillas ($p=0$). 
As shown in Fig. \ref{f:Farina_nonMarkov}, the non-Markovianity measure is a monotonously increasing function of $g/\kappa$ and it is equal to zero under a certain threshold; we thus observe that, also in this case, in the regime where one obtains the maximum ergotropy (the large loss limit, $g/\kappa \to 0$), the effective dynamics of the battery subsystem is in fact Markovian.

\section{Conclusions}
\label{sec:conclusions}
In this work, we studied the discrete-time and continuous-time dynamics of a quantum battery interacting with a collisional model that acts as the environment. Specifically, we described how a discrete-time collisional model is able to mimic  
an environment with memory by introducing an ancilla-ancilla partial swap interaction with a certain probability $p$. 
It was therefore possible to explore the transition between a memoryless environment (obtained for $p=0$) and an environment able to induce strong memory effects  on the battery and the charger 
(for values of $p$ approaching $1$), while observing the properties of the corresponding steady states of the dynamics. 
In particular, we observed a non-monotonic behaviour of the ergotropy as a function of the swap probability $p$, and the presence of a maximum for a large value of $p$, smaller than the limiting value $p=1$. 
As we also observed that energy decreases monotonically with $p$, we came to the conclusion that the ergotropy behaviour is mainly driven by the change in the steady-state purity.

We have also found that the maximum ergotropy achievable at steady state in all the different parameter regimes corresponds to the same value that was obtained with a Markovian environment in the strong dissipative regime, 
which suggests that the maximum ergotropy is fixed by the operatorial properties of the interaction, regardless of the features of the environment. 
On the other hand, the presence of memory effects induced by the environment 
allows us to approach the maximum value of ergotropy in a broader region of the parameter space,
and in a shorter amount of time, thus boosting the charging speed of the battery.

The behaviour observed in the discrete-time model is consistent with what we observe in the regime of parameters where the continuous-time limit exists, that is, for $p\approx 1$. Specifically, we observe that the ergotropy increases by increasing the memory rate $\Gamma$ characterizing this dynamics, and thus by decreasing the 
capability of the environment to store information about the evolution of the system.

By using a definite measure of non-Markovianity, the geometrical measure, we investigated 
the amount of memory effects affecting the battery and the charge seen as a joint system,
as well as the battery only. While the non-Markovianity of the joint system
shows the expected monotonically increasing behavior as a function of the ancilla-ancilla collision
parameter $p$, the non-Markovianity of the battery has a non-monotonic behavior as a function of $p$,
possibly including regions where there are actually no memory effects, as quantified by the geometrical measure of non-Markovianity. Remarkably, the maximum of the steady state ergotropy lies in regions of parameters where the battery dynamics is Markovian, as the combined effects of the memory due to, respectively, the direct coherent interaction with the charger and the backflow of information mediated by the ancilla-ancilla collisions cancel each other.

Our work is one of the first attempts in understanding the role of memory effects in quantum battery charging and is particularly suitable for a direct experimental implementation in the next future on actual quantum devices, as quantum collisional models have recently been demonstrated on different quantum simulation platforms \cite{Cuevas2019,GarciaPerez2020,cattaneo2022}. We believe that our study paves the way to further research in this direction: it will be interesting and necessary to understand if the behaviour we have observed will be confirmed when considering quantum batteries with larger dimensionality, that is, going beyond the qubit case, and/or for other noise-models characterized by different forms of the coupling or by different sources of non-Markovianity. In the first instance, one could for example look at dephasing, where  non-trivial thermodynamic behaviour has recently been found~\cite{Popovic2021}. Concerning non-Markovianity, one could consider collisional models exhibiting information backflow with or without system-environment correlations~\cite{McCloskey2014}: it has been indeed shown how the two scenarios have an impact on thermodynamic properties, such as the entropy production rate~\cite{Senyasa2022}. These results suggest that it may be interesting to exploit these models in order to further study the actual relationship between ergotropy, decoherence, information backflow and system-environment correlations.
\\
\begin{acknowledgments}
DM acknowledges financial support from MUR under the “PON Ricerca e Innovazione 2014-2020”.
MACR acknowledges financial support from the Academy of Finland via the Centre of Excellence program (Project No. 336810).
MGG and AS acknowledge support from UniMi via PSR-2 2020 and PSR-2 2021.
The computer resources of the Finnish IT Center for Science (CSC) and the FGCI project (Finland) are acknowledged.
\end{acknowledgments}

\bibliography{reference}
\end{document}